\documentclass[manuscript]{aastex}

\usepackage{natbib}

\slugcomment{Accepted for publication in ApJ, March 21, 2016}

\shorttitle{Multiple current sheet systems}
\shortauthors{Burgess et al.}

\begin{document}

\title{Multiple current sheet systems in the outer heliosphere: Energy release and turbulence}

\author{D. Burgess, P. W. Gingell}
\affil{School of Physics and Astronomy, Queen Mary University of London, London E1 4NS, UK}

\and

\author{L. Matteini}
\affil{Imperial College London, London SW7 2AZ, UK}

\begin{abstract}
In the outer heliosphere, beyond the solar wind termination shock, it is expected that the warped heliospheric current sheet forms a region of closely-packed, multiple, thin current sheets. Such a system may be subject to the ion-kinetic tearing instability, and hence generate magnetic islands and hot populations of ions associated with magnetic reconnection. Reconnection processes in this environment have important implications for local particle transport, and for particle acceleration at reconnection sites and in turbulence. We study this complex environment by means of three-dimensional hybrid simulations over long time scales, in order to capture the evolution from linear growth of the tearing instability to a fully developed turbulent state at late times. The final state develops from the highly ordered initial state via both forward and inverse cascades. Component and spectral anisotropy in the magnetic fluctuations is present when a guide field is included. The inclusion of a population of new-born interstellar pickup protons does not strongly affect these results. Finally, we conclude that reconnection between multiple current sheets can act as an important source of turbulence in the outer heliosphere, with implications for energetic particle acceleration and propagation.

\end{abstract}


\section{Introduction}

Prior to Voyager crossing the heliospheric termination shock (Voyager 1 in 2004, Voyager 2 in 2007) it was widely accepted that the anomalous cosmic ray (ACR) component comprised interstellar pickup ions accelerated at the solar wind termination shock in the outer heliosphere. The ACR component is a part of the energetic particle spectrum at total energies of 10 -- 100 MeV, solar modulated, and predominantly single-charged, which is an enhancement over the galactic cosmic ray spectrum. However, observations of the ACR component made at the termination shock \citep{stone:2005,stone:2008} disagreed with the expectations of a straightforward application of shock acceleration theory, in that the fluxes did not peak at the shock crossing, and there was not a smooth power law energy spectrum at and downstream of the shock. The discrepancy between the Voyager observations and the initial expectations of diffusive shock acceleration (DSA) theory has led to a re-examination of shock acceleration theory and the suggestion of new mechanisms for the acceleration of the ACR component, as reviewed by \citet{giacalone:2012}. On the other hand, shock acceleration theories have been refined by considering the effects of a non-spherical symmetric termination shocks \citep{mccomas:2006,kota:2008,jokipii:2013}, and recently it has been suggested that solar cycle changes in the solar wind should be taken into account in order to interpret the Voyager observations within the framework of shock acceleration theory \citep{senanayake:2015}. Other recent work refining the model of acceleration at the termination shock includes the combination of diffusive shock acceleration and acceleration via interacting flux ropes behind the termination shock \citep{zank:2015}. This work, based on kinetic transport theories developed to model energetic particle propagation in the presence of evolving flux ropes \citep{zank:2014,leroux:2015}, puts forward an explanation of why the observed energetic particle spectra are harder than predicted by DSA alone, and why fluxes peak downstream of the termination shock.

Alternatives to shock acceleration at the termination shock include diffusive compression acceleration due to plasma compressions in the heliosheath \citep{fisk:2009} and magnetic reconnection in the heliosheath. We will consider the latter in more detail. The misalignment of the rotational and magnetic axes of the Sun produces a region in the distant heliosphere where the dominant azimuthal magnetic field periodically reverses its polarity due to the warped heliospheric current sheet (HCS). It was recognized that as the solar wind compresses at the termination shock, and slows towards the heliopause, this region of sectors with alternating magnetic field direction will compress, causing thinning of the  heliospheric current sheet and possibly leading to magnetic reconnection \citep{lazarian:2009,drake:2010}. \citet{lazarian:2009} proposed a model for the acceleration of the ACR component based on turbulent diffusive acceleration where the turbulence is driven by reconnection on the large scale, and a first order Fermi process operates by energetic particles interacting with reconnection outflows with contracting magnetic field line loops.

Motivated by the concept of a sectored heliosheath, \citet{drake:2010} considered the evolution of a system of multiple current sheets in a two-dimensional (2-D) geometry using full-particle particle-in-cell (PIC) simulations. High energy tails were found to form on the electron and ion distribution functions, and acceleration due to contraction of magnetic islands has found to be responsible for the energy gain of the most energetic ions.  The low $\beta$ plasma ($\beta_e=0.083$) used by \citet{drake:2010} favors a scenario where particle energization is dominated by energy release from the magnetic field. For higher beta, up to $\beta=4.8$, it was found, again for 2-D PIC simulations, that reconnection was quenched by the plasma reaching a marginal firehose  unstable condition, so that the end state (at least over the duration of the simulation) consists of extended magnetic islands \citep{schoeffler:2011}. It has also been proposed that the sectored heliosheath, consisting of multiple current sheets that reconnect due to compression, is important for the ``porosity'' of the heliopause, creating a multiple island magnetic field structure which leads to patchy reconnection at the heliopause \citep{swisdak:2013}. This was suggested to explain the temporal sequence of ACR and galactic cosmic rays seen at the Voyager~1 crossing of the heliopause \citep{krimigis:2013}. The question of the role of sector structure in the heliosheath is also important for cosmic ray propagation \citep{florinski:2012}.

In this paper we explore the question of energy release and particle acceleration in multiple current sheet (MCS) systems, using simulations which include several effects not modeled in earlier 2-D full particle PIC simulations. We concentrate on ion dynamics in the system by using a hybrid simulation with the electron response modeled as a fluid, and include a newly-born pickup ion distribution. The hybrid simulation method has proved itself to be a useful and reliable tool for studying solar wind turbulence \citep[e.g.,][]{franci:2015b}, indicating that for dynamics on scales larger than those of the electrons, the solar wind behavior is well-captured. The current sheets are initialized in a force-free configuration with a rotation of the magnetic field direction. We also include, for comparison and more realistically, the effects of a guide field. Finally, the simulation is fully three-dimensional (3-D), capturing the dynamics of reconnection which are known to depend on both ion tearing and drift-kink instabilities, which combine to produce three-dimensional structuring \citep[e.g.,][]{daughton:2011,gingell:2015}. Of course, computational constraints do limit the realism of the simulations: the spatial domain is  small compared to the scale of the heliosheath; the separation between the current sheets is relatively small so that the time scale for interaction between them can be covered in the simulation. The separation of the current sheets in the simulation corresponds to approximately $5\times 10^4$km, which is much less than the observed scale of sector structures in the heliosheath, but only slightly smaller than realistic for fine-scale structure within the HCS. If the separation in the simulation were increased, the time scale for interaction between adjacent current sheets would also increase, so that not only would the simulation have to be larger, but also run longer. Nevertheless, the process of interaction between the current sheets would be similar, since, as described later, it proceeds via both forward and inverse cascade in wave vector space.

The simulation results indicate that the highly symmetric and ordered initial state of a MCS system evolves to a complex three-dimensional state, with energy release and energization of both solar wind (SW) and pickup ion (PUI) distributions. But, perhaps most interestingly, the long time end state of our simulations appears, by different measures, to become close to well-developed turbulence. A power law form of the power spectrum of magnetic fluctuations develops due to both inverse and forward cascade. The turbulence created by the energy release from the MCS system is anisotropic for the case where a guide field is present, as found for solar wind turbulence. Below we discuss the arrangement and parameters of the simulations, present results and conclude with some remarks on the role of MCS systems in the outer heliosphere.

\section{Simulations}

We investigate the evolution of a multiple current sheet system using the three-dimensional hybrid simulation code HYPSI, previously used to study the tearing and drift-kink instabilities for current sheets in the solar wind \citep{gingell:2015}. The hybrid plasma simulation method combines a kinetic, particle-in-cell treatment of the ion species with a charge-neutralizing, massless and adiabatic electron fluid. Maxwell's equations are are solved in the low-frequency Darwin limit with zero resistivity, using the CAM-CL method described by \citet{matthews:1994}.

The simulation uses a grid of $(N_x,N_y,N_z)=(120,120,120)$ cells, with a resolution $\Delta x = \Delta y = \Delta z = 0.5 d_\mathrm{i}$, where $d_\mathrm{i}=v_\mathrm{A}/\Omega_\mathrm{i}$ is the ion inertial length. The total simulation domain size is $(60 d_\mathrm{i})^3$. Distance and time are normalized to units of the ion inertial length $d_\mathrm{i}$ and inverse ion gyrofrequency $t_\Omega = \Omega^{-1}_\mathrm{i}$, respectively; velocity is normalized to the Alfv\'en speed $v_\mathrm{A}$.

The simulation is initialized with eight current sheets parallel to the $y$ - $z$ plane, with a spacing of 7.5 $d_\mathrm{i}$. At each current sheet the $|B_y| = B_0$ component of the magnetic field reverses over a length scale $L = d_\mathrm{i}$. For some simulations a uniform guide field in the $z$ direction is added, such that initially $B_z=B_y$ and the background field strength $B^2_\mathrm{g}=B_y^2+B_z^2 = B_0^2$ is preserved. For most of the simulations we discuss the current sheets are initialized in a ``force-free'' configuration, which consists of a rotation of the magnetic field in the $y$-$z$ plane as the current sheet is crossed in the $x$ direction. This configuration is more typical of current sheets in the solar wind,  in contrast to the Harris current sheet equilibrium used in the full particle simulations reported by \citet{drake:2010}.

The simulation includes a uniform SW proton population with a Maxwellian velocity distribution, with plasma beta $\beta_\mathrm{sw}=0.5$, together with a population representing recently created interstellar pickup ions. The pickup ion population has a density $n_\mathrm{pui}$ of 10\% of the background solar wind, and a velocity distribution with the form of a spherical shell, centred on the solar wind velocity and with a finite width. The characteristic shell velocity is $4 v_\mathrm{A}$ and the velocity spread of the shell is $0.7 v_\mathrm{A}$. The assumption of a spherical shell in velocity space for the pickup ion velocity distribution takes into account rapid pitch angle scattering which will isotropize the distribution function. It does not take into account pickup ions at lower or high energy which are ionized closer to the Sun and then convected outwards to the heliosheath, undergoing diffusive transport in the turbulent solar wind. There are no direct measurements of the low energy pickup ions in the outer heliosphere, so there is some uncertainty about their distribution and density. Other studies have used higher densities and different pickup velocity distributions. Our choice of $\beta_\mathrm{sw}$ is appropriate for the heliosheath \citep{richardson:2011,burlaga:2009}. We discuss the effects of parameter variations later.

\section{Results}

\begin{figure}
\begin{tabular}{cc}
(a) &\\[-1ex]
 & \includegraphics[width=16cm]{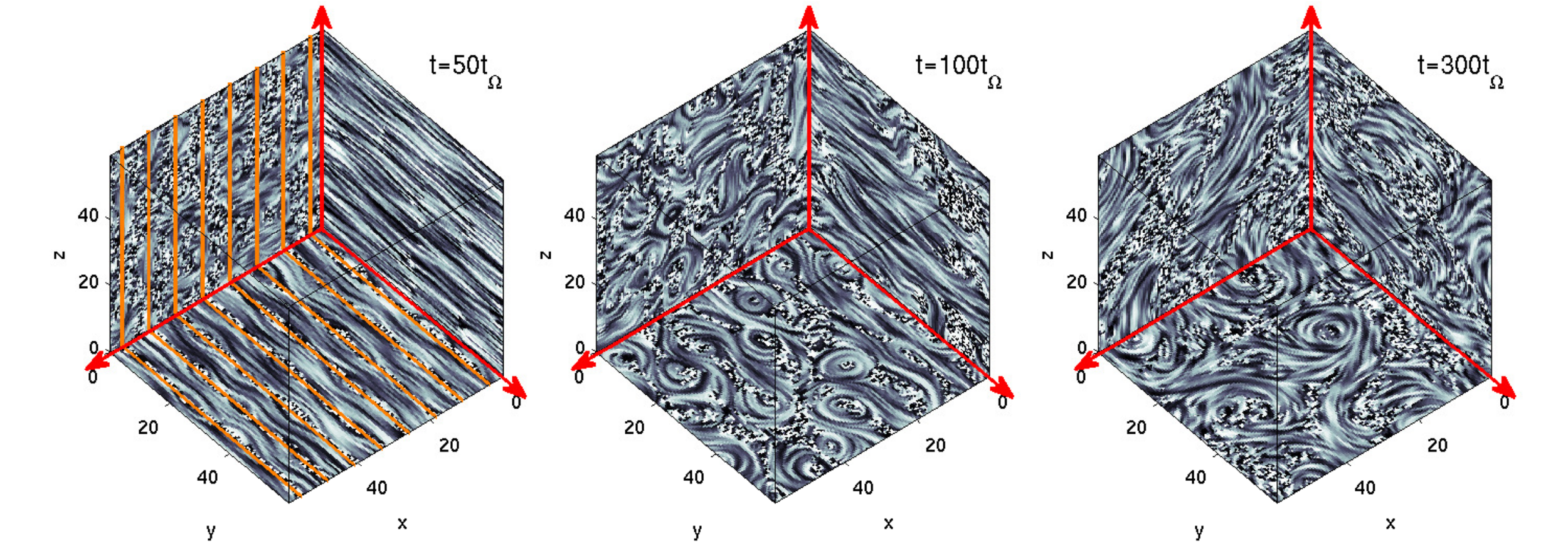} \\
(b) &\\[-1ex]
& \includegraphics[width=16cm]{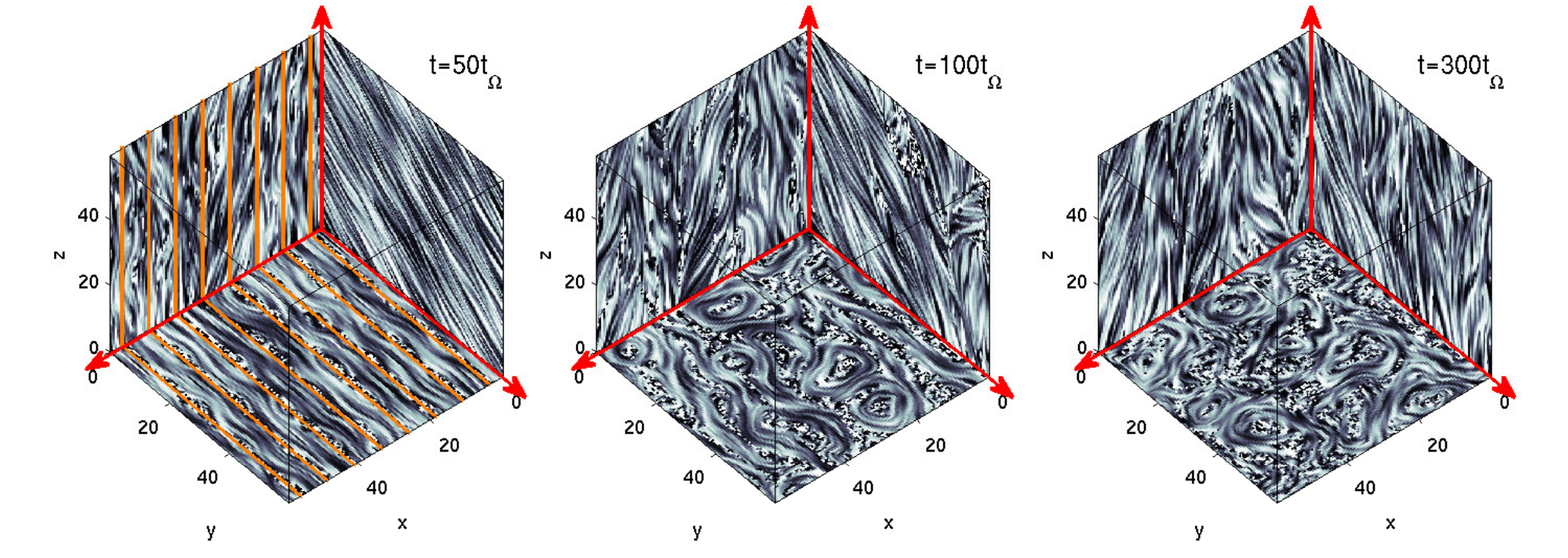}
\end{tabular}
\caption{Evolution of multiple current sheet system at times $\Omega_\mathrm{i} t=50, 100, 300$ with (a) no guide field, and (b) a guide field of $B_\mathrm{g}=1$, using a line integral convolution (LIC) representation of the magnetic field lines on three orthogonal planes. (The plots are produced in MATLAB with routines from Toolbox image.) The configuration of the initial current sheets is shown in the left-hand panels with orange lines.}
\label{fig:1}
\end{figure}

The time evolution of the three-dimensional simulation is illustrated in Figure~\ref{fig:1}, which shows a line integral convolution (LIC) visualization of the projection of the magnetic field in each of the three cartesian coordinate planes. LIC provides a texture-based flow visualization which gives an indication of the field line topology but not its polarity (i.e., direction). It provides a dense representation without being as visually cluttered as following multiple field lines.

For the case of zero guide field the simulations display the following general pattern of evolution: The initial state of multiple planar current sheets is symmetric and highly ordered, as indicated by the orange lines shown in Figure~1a which are the projections of the current sheets onto the coordinate planes. Initially the magnetic field is directed in alternately the $+y$ and $-y$ directions, so that all field lines are parallel to the $x-y$ plane. In the early stages the thin current sheets begin to reconnect, forming approximately two-dimensional structured magnetic islands, with field lines remaining roughly in their original $x-y$ planes. This can be seen in Figure~1a at $t=50t_\Omega$ from the $y$~oriented structuring in the $y-z$ plane, and the wavy ``island-like'' modulation of the structure in the $x-y$ plane. At this stage the initial 8-fold translational symmetry of the initial conditions is still visible.

As time progresses and reconnection continues, the magnetic islands continue to grow in size, begin to interact and take on structuring perpendicular to the original reconnection in the $x-y$ plane. This can be seen in Figure~\ref{fig:1}a at $t=100t_\Omega$ where the islands from adjacent current sheets have grown large enough to interact  although there is still a remnant of the initial 8-fold symmetry (seen in $x-y$ plane), and the islands are now structured in the $x-z$ plane. There is still predominantly $y$ directed structuring in the $y-z$ plane, as a remnant of the initial conditions.  By the end of the simulation (Figure~\ref{fig:1}a $t=300t_\Omega$) the system has reached a state with a well-developed three-dimensional structure in the magnetic field, with little difference in character between the $x-z$ and $y-z$ planes. There is still some indication of magnetic island structuring (or helical structuring) visible in the $x-y$ plane, but the characteristic size of the islands (or vortices) has become larger than the initial current sheet separation. Overall the evolution of the simulation can be summarized as a transition from a highly symmetric and ordered system to one with complex, three-dimensional structure.

The effect on the evolution of the addition of a guide field $B_\mathrm{g}=1$ is shown in  Figure~\ref{fig:1}b. The reconnecting magnetic fields are now angled to the $x-y$ plane at $\pm 45^\circ$, and indicated by the linear structuring still evident in the $y-z$ plane at $t=50t_\Omega$. The early stage evolution is visible in the $x-y$ plane and is similar to the zero guide field case, with the formation of initially 2-D islands which grow and then begin to interact. However, the guide field clearly introduces anisotropy associated with the mean magnetic field direction (i.e., the $z$ direction), visible in the linear structuring in the $x-z$ and $y-z$ planes. At late times in the simulation, from the $x-y$ plane, it is seen that there are still remnants of island-line structuring, with a range of sizes including larger than the initial current sheet separation.

Quantifying reconnection in three-dimensions can be computationally and conceptually difficult \citep{schindler:1988,yeates:2011}, so we use a simple metric, or proxy for reconnected flux which allows us to compare a number of different simulations. The time evolution of $\int |B_x| d^3x$, over the full simulation domain is shown in Figure~2a. Comparing the results for the force-free current sheet initialization with (blue curve) and without (black curve) the PUI distribution shows that there are some small differences initially, but after about $t=50t_\Omega$ the differences are negligible. The differences at early times might be attributed to the initial particle noise. In both cases 200 particles per cell are used, but for the case with PUI there is an equal split between the PUI and SW populations, so that the initial particle noise is higher for this case. The addition of a guide field (green curve without PUI, cyan curve with PUI) leads to a lower level of reconnected flux at late times, consistent with reconnection of the in-plane anti-parallel component. (The magnetic field magnitude is the same for both zero and non-zero guide field, so in the guide field case, the component of the field which is anti-parallel across the current sheet is smaller; consequently, the amount of flux initially available for reconnection is also smaller.) At intermediate times the growth rate appears slower than for the zero guide field case, and at very early times a similar discrepancy due to initial particle noise is visible.

For comparison we have also carried out simulations in which the current sheets are initially in a Harris equilibrium, with the PUI population included as part of the uniform background (Figure~2a red curve). Some relatively minor differences are seen. At very early times the growth rate is similar to the force-free case with guide field and PUI, perhaps indicating that ion-tearing dominates at this time. However, the long term evolution shows a slower growth rate and a lower level at late times, although there is some indication that it has not yet stopped increasing. These results indicate that the growth of islands and their interaction, after an initial stage of fast growth, is slower in the nonlinear stage for the Harris sheet initialization compared with the force-free case. This is consistent with previous results demonstrating reduced reconnection rates for Harris sheets subject to a drift-kink instability \citep{gingell:2015}.

The effects of reconnection on particle energization is demonstrated in Figure~2b which shows the time evolution of the total kinetic energy of the solar wind component as a ratio to its initial value. Again, similar behavior is seen for the $B_\mathrm{g}=0$ case for both with and without PUI. The overall behavior for the guide field case $B_\mathrm{g}=1$ (both with and without PUI) is similar but growth in the kinetic energy is slower and there is a lower saturation level. Figure~3a,b shows the particle energy distributions for SW and PUI populations at $t=0$ and $t=300t_\Omega$, for the zero guide field case. It is evident that both distributions show heating. The heating for the solar wind component is relatively modest, as expected from the relatively high initial plasma beta. The PUI distribution shows stronger heating and broadening; examination of the $v_\| - v_\perp$ distribution shows a broadening of the initial shell to both higher and lower energies, with the overall distribution remaining isotropic on average.


We examine the transport of ion species $s$ by considering the time dependence of the cross-boundary, one-dimensional mean-square displacement of the simulation particles $\left<\Delta x_s^2 \right>$. Here, the mean-square displacement is taken in the $x$-direction, i.e., parallel to the initial current sheet normals, representing the radial direction in the heliospheric coordinate system. In this way, we measure particle transport in the context of porosity of the region across the current sheets. Particle transport can be characterised by a power law increase in mean-square displacement, such that
\begin{equation}
\left<\Delta x_s^2\right> \propto \left(\Omega_s t \right)^\gamma.
\label{eqn:msd_powerlaw}
\end{equation}
The case $\gamma = 1$ represents classical diffusion, $\gamma < 1$ represents a subdiffusive regime, $\gamma > 1$ represents a hyper-diffusive regime, and $\gamma = 2$ represents ballistic transport \citep{metzler:2000,zimbardo:2013}.

Using this method, anomalous particle diffusion has been demonstrated in several contexts for space, astrophysical and laboratory plasmas. For example, cross-field hyper-diffusion of ions, with $\gamma > 2$,  has been demonstrated to occur as a result of the growth of non-linear instabilities at planetary and tokamak boundary layers \citep[e.g.,][]{cowee:2009,gingell:2015}, and in the context of diffusive shock acceleration \citep{giacalone:2013}. Turbulence has also been shown to play an important role in hyper-diffusive transport both parallel \citep{zimbardo:2006,pommois:2007} and perpendicular \citep{gustafson:2012} to the local magnetic field direction. However, we note that the transport exponent $\gamma$ can be strongly dependent on plasma beta, kinetic scale lengths and the isotropy, amplitude and time-dependence of turbulent fluctuations. Indeed, sub-diffusive and classically diffusive regimes have also been observed in many of these systems.

The mean-square displacement is shown for our simulations in Figure~3c for SW (red) and PUI (blue) species. Dashed lines represent diffusive power law $\gamma = 1$ and hyper-diffusive power law $\gamma = 2$, overlaid on plots for PUI and SW species respectively. This demonstrates that the heating of the PUI population seen in Figure~3b does not result in faster cross-boundary transport, as observed for the lower energy SW population. This is consistent with cross-boundary hyperdiffusivity occurring due to reconnection generating cross-boundary magnetic flux at small scales. In this case, the PUI population is decoupled from these small-scale changes in topology as a result of its relatively low density and large gyro-radius.


Since the magnetic field appears to develop a complex three-dimensional structure over the course of the simulation, it is reasonable to consider its turbulence properties. By the end of the simulation at $t=300t_\Omega$ the magnetic field fluctuations have a power spectrum with a power law over a large section of the available wave vector range, indicating that a turbulent description of the final state might be appropriate. Figure~4a shows the trace power spectrum of magnetic field fluctuations for the zero guide case, which at the end of the simulation approximately follows a power law $P(|k|) \sim |k| ^{-7/3}$. ($P_\mathrm{tot}=\int P(|k|)\,dk$ gives the total magnetic energy.) This result is obtained for the case of initialization from force-free current sheets, both with (green curve) and without PUI (red curve). For comparison the power spectrum at $t=0$ is also plotted (blue curve), showing the enhancements at regular spacing in $|k|$ corresponding to spatial harmonics of the inter-current sheet spacing. The evolution of the spectrum from $t=0$ to later times consists of infilling between the initial peaks and towards small $k$, reduction of the maxima, and development of a stationary power law. If the final spectrum evolved from inertial MHD turbulence, one might expect a power law slope of $-5/3$. The slightly steeper slope found in the simulation could be a consequence of a combination of the initial driving, the effects of inverse cascade with a limited range of available wavevector space at small $k$, and kinetic effects. The rising part of the spectrum at high $k$ is due to the effects of finite particle number noise, and the definition of $P(|k|)$. There might also be a contribution from the remnant of the initial driving at high $k$.

It has been noted earlier that the inclusion of a guide field introduces an anisotropy in the spatial structure of the magnetic field, and this is also seen in the power spectra of the magnetic field components shown in Figure~4b. The mean field in this case, for $B_\mathrm{g}=1$, is in the $z$ direction, and the power spectrum for $B_z$ (as well as that for the total magnetic field) follows an approximate power law $\sim |k| ^{-1}$, whereas the power spectra for the $x$ and $y$ (i.e., perpendicular) components are at a higher level and follow a power law $\sim |k| ^{-7/3}$. The insets in Figure~4 show the total power spectra as a function of $(k_\parallel,k_\perp)$, illustrating the effect of a guide field on the spectral anisotropy. The simulation with guide field shows clear component and spectral anisotropy, similar to that observed in the solar wind \citep[e.g.,][]{wicks:2010,chen:2010}.

\section{Ion Heating}

In this section, we will discuss in more detail the sources of particle heating seen in Figure~3. First we note that our simulations are of a relaxation process, without continual driving which would produce greater heating. Secondly, we note that, given the difference in the initial energies of the SW and PUI populations, the energy release to the PUI population results in much smaller gain in energy relative to the initial total. Hence, changes in the PUI distribution function, such as growth or thermalisation of temperature anisotropies, are much less significant than for the SW population. For this reason, we present signatures of particle heating in the solar wind population only. 

We determine the period at which strong heating occurs by examining the solar wind ion energisation parallel and perpendicular to the local magnetic field (Figure~\ref{fig:5}). From the time evolution of the fraction of ions above a given energy threshold, we can identify two periods of significant ion heating: i) for $t\Omega_\mathrm{i} < 40$ we observe growth of the perpendicular energy, and decay of the parallel energy; ii) for $40 < t\Omega_\mathrm{i} < 150$ we observe slower growth of perpendicular energy, and growth of the parallel energy. These two periods correspond to the linear phase of the tearing instability, and the non-linear phase during which adjacent magnetic islands begin to merge. This suggests that the principle mechanism of particle energisation is magnetic reconnection, in agreement with previous investigations of ion heating at ion-scale current sheets \citep{gingell:2015}. At late times ($t\Omega_\mathrm{i} > 150$), when the reconnection rate is low and the turbulent spectrum is fully developed, the ion energisation rate is small compared to the preceding periods. This suggests that turbulent heating via second order Fermi processes does not significantly contribute to the observed heating. However, the rise in both parallel and perpendicular ion energies during the turbulent phase is consistent with rise of $T_\perp$ and $T_\parallel$ in high-resolution 2-D simulations of ion scale turbulence \citep{franci:2015b}.

To further characterise the heating processes occurring in this multiple current sheet system, we can also determine the location of ion energisation using maps of local temperature anisotropy. In the top row of Figure~\ref{fig:6}, the temperature anisotropy of the solar wind ion population $T_{\perp}/T_{\parallel}$ is displayed in an $x$-$y$ plane, at $t\Omega_\mathrm{i} = 40$, representative of the first period of heating associated with the linear tearing mode, and at $t\Omega_\mathrm{i} = 100$, for the second period of heating associated with non-linear tearing (i.e. island merging). During the linear phase of the tearing evolution, we find high anisotropy ($T_\perp > T_{\parallel}$ at the x-points (dark blue). The high perpendicular temperature at the x-points is consistent with ions following Speiser-like trajectories \citep{hietala:2015}. During the non-linear phase at $t\Omega_\mathrm{i} = 100$, we also find a clear parallel temperature excess (red) along the separatrices of the magnetic islands, adjacent to the x-points. This is perhaps most clear at $(x,y) = (15,45)d_\mathrm{i}$. These high parallel temperature structures are seen to be associated with reconnection exhausts, consistent with observations of reconnection exhausts and 2-D simulations \citep{hietala:2015,drake:2010}. We are therefore able to confirm that this model of reconnection exhausts persists in a fully 3-D environment.

The middle row of Figure~\ref{fig:6} shows highlighted regions where the solar wind ion population is stable (blue) and unstable (red) to the firehose instability, with instability condition $(\beta_{\parallel}-\beta_{\perp}) > 1$. At early times (left), regions of local instability are uncommon. However, their proximity to x-points supports the association of high $T_\parallel$ regions with reconnection exhausts. At $t\Omega_\mathrm{i} = 100$ (right), regions of local instability are both more common and of larger size. Although the resolution of the simulation is insufficient to directly and clearly display firehose wave modes, we note that the greater occurrence of firehose unstable regions at later times is concurrent with the isotropisation of the solar wind ion temperature. This suggests that firehose scattering may play a role in the isotropisation, as observed by \citet{hellinger:2015}. Furthermore, where temperature structures at reconnection sites are sufficiently resolved, the location of firehose unstable regions is consistent with observations and simulations \citep{hietala:2015}, i.e., enhancement of $T_\perp$ at  along the neutral line, enhancement of $T_\parallel$ at the edges of the magnetic island, and a firehose unstable region in-between. As before, we have therefore validated the results of earlier 2-D simulations in a fully 3-D system. We also observe regions with very high instability parameter $(\beta_{\parallel}-\beta_{\perp})  > 4$. As discussed by \citet{hietala:2015}, this may be a consequence of a driving rate much faster than the growth rate of the instability, caused by continuous refilling of the unstable regions at the boundaries of the reconnection exhausts.

\section{Conclusions}

With three-dimensional hybrid simulations we have studied the evolution of MCS systems in the heliosheath, including interstellar pickup ions, to examine what happens when the current sheets thin, reconnect and are close enough to interact. This situation has been suggested as important for the acceleration of the ACR component \citep{drake:2010,schoeffler:2011} and for driving patchy reconnection at the heliopause \citep{swisdak:2013}. Reconnection via the ion-tearing instability dominates at early time. Subsequent interaction between islands on adjacent islands leads to the development of complex three-dimensional structures in the magnetic field, and some energization of the particle populations. Inclusion of a guide field produces anisotropy in the magnetic field fluctuations, with slower reconnection and energization, at all stages in the simulation. The presence of a pickup ion population, at the level of 10\% of the solar wind density, does not have any strong effect on reconnection rates or the overall evolution, although there is some evidence of differences in the transport properties of the pickup and solar wind ion populations.

In all the simulations, the late stage behavior appears turbulent, with power law behavior in the magnetic fluctuation power spectra, and, for the case of a guide field, with component and spectral anisotropy. The system thus exhibits both forward and inverse cascade with a transfer of magnetic energy from a highly ordered initial state to one with fluctuations across all available wave vector space. It is interesting to note that the multiple force-free current sheet system studied here is similar to the so-called sheet-pinch configuration studied with a three-dimensional full-particle PIC simulation \citep{bowers:2007}, which also showed energy release and spectral energy transfer via both forward and inverse cascades. Thus, although in our simulations we do not see dramatic energization, as might be used to explain acceleration of the ACR component, we would infer that the folded (or ``pleated'') MCS system of the heliospheric current sheet will be a driver of turbulence in the heliosheath, and could contribute to the propagation properties of the ACR component within the heliosheath. Obviously this additional driver of turbulence could only operate in limited range of heliolatitude corresponding to the warping of the HCS. The question of a porous heliopause could then be addressed by investigating the consequences of the turbulence for the structure of the heliopause.

We have carried out a number of simulations with different parameters for the SW and PUI populations, including higher density (20\%) and energy (shell velocity $10 v_\mathrm{A}$) for the PUI and lower $\beta_\mathrm{sw}=0.1$. Broadly speaking, the final magnetic power spectra are unchanged, showing the same power law behavior. The energization of the SW population is roughly the same, except that the heating rate is higher for low $\beta_\mathrm{sw}$. The reconnection rates are similar for higher PUI density and energy, but with the onset faster due to enhancement of the drift-kink instability \citep{gingell:2015}, and the late-time rate slower due to broadening of the current sheet. The solar wind ion distribution has been assumed to be an isotropic Maxwellian. The inclusion of a temperature anisotropy would affect the reconnection rate in the current sheets, for example, due to the role of ion cyclotron or firehose instabilities, depending on whether $T_\perp>T_\|$ or $T_\perp<T_\|$, respectively \citep{matteini:2013,gingell:2015}. The effect of using a kappa solar wind distribution on the evolution of the current sheets remains a topic for future work.

The study of energy release in MCS systems has a number of important implications. Firstly, there are other environments where such a structured, sectored magnetic field might occur, either in the solar corona, or within the heliospheric current sheet itself. For example, \citet{vlahos:2004} proposed a model for particle acceleration in solar flares with a fragmented, localized system of regions of enhanced electric field that might be produced in a multiple current sheet system. Secondly, the transition from a ordered magnetic field to a turbulent state suggests that it can be a method of initializing turbulence in simulations, in order to study the properties of that turbulence. As such it could serve as a way to initialize turbulence without making assumptions about the fluctuations, which is a characteristic of the methods commonly used for simulations of decaying turbulence. It also makes a useful comparison with other methods of multi-scale driving of turbulence \citep{yoo:2014}. It is an important question to what extent the properties of turbulence depend on how it is driven, and further study of the turbulence produced by the MCS system is currently in progress.

\acknowledgments

The authors thank Joe Giacalone for insightful comments.
This work was supported by the UK Science and Technology Facilities Council (STFC) grants ST/J001546/1 and ST/K001051/1. The research leading to the presented results has received funding from the European Commission's Seventh Framework Programme FP7 under the grant agreement SHOCK (project number 284515). This work used the DiRAC Complexity system, operated by the University of Leicester IT Services, which forms part of the STFC DiRAC HPC Facility (www.dirac.ac.uk). This equipment is funded by BIS National EInfrastructure capital grant ST/K000373/1 and STFC DiRAC Operations grant ST/K0003259/1. DiRAC is part of the National E-Infrastructure.


\begin{thebibliography}{}
\expandafter\ifx\csname natexlab\endcsname\relax\def\natexlab#1{#1}\fi

\bibitem[{{Bowers} \& {Li}(2007)}]{bowers:2007}
{Bowers}, K., \& {Li}, H. 2007, Phys. Rev. Lett., 98, 035002

\bibitem[{{Burlaga} \& {Ness}(2009)}]{burlaga:2009}
{Burlaga}, L.~F., \& {Ness}, N.~F. 2009, \apj, 703, 311

\bibitem[{{Chen} {et~al.}(2010){Chen}, {Horbury}, {Schekochihin}, {Wicks},
  {Alexandrova}, \& {Mitchell}}]{chen:2010}
{Chen}, C.~H.~K., {Horbury}, T.~S., {Schekochihin}, A.~A., {et~al.} 2010, Phys.
  Rev. Lett., 104, 255002

\bibitem[{{Cowee} {et~al.}(2009){Cowee}, {Winske}, \& {Gary}}]{cowee:2009}
{Cowee}, M.~M., {Winske}, D., \& {Gary}, S.~P. 2009, J. Geophys. Res., 114,
  A10209

\bibitem[{{Daughton} {et~al.}(2011){Daughton}, {Roytershteyn}, {Karimabadi},
  {Yin}, {Albright}, {Bergen}, \& {Bowers}}]{daughton:2011}
{Daughton}, W., {Roytershteyn}, V., {Karimabadi}, H., {et~al.} 2011, Nature
  Physics, 7, 539

\bibitem[{{Drake} {et~al.}(2010){Drake}, {Opher}, {Swisdak}, \&
  {Chamoun}}]{drake:2010}
{Drake}, J.~F., {Opher}, M., {Swisdak}, M., \& {Chamoun}, J.~N. 2010, \apj,
  709, 963

\bibitem[{{Fisk} \& {Gloeckler}(2009)}]{fisk:2009}
{Fisk}, L.~A., \& {Gloeckler}, G. 2009, Advances in Space Research, 43, 1471

\bibitem[{{Florinski} {et~al.}(2012){Florinski}, {Alouani-Bibi}, {Kota}, \&
  {Guo}}]{florinski:2012}
{Florinski}, V., {Alouani-Bibi}, F., {Kota}, J., \& {Guo}, X. 2012, \apj, 754,
  31

\bibitem[{{Franci} {et~al.}(2015){Franci}, {Landi}, {Matteini}, {Verdini}, \&
  {Hellinger}}]{franci:2015b}
{Franci}, L., {Landi}, S., {Matteini}, L., {Verdini}, A., \& {Hellinger}, P.
  2015, \apj, 812, 21

\bibitem[{{Giacalone}(2013)}]{giacalone:2013}
{Giacalone}, J. 2013, \ssr, 176, 73

\bibitem[{{Giacalone} {et~al.}(2012){Giacalone}, {Drake}, \&
  {Jokipii}}]{giacalone:2012}
{Giacalone}, J., {Drake}, J.~F., \& {Jokipii}, J.~R. 2012, \ssr, 173, 283

\bibitem[{{Gingell} {et~al.}(2015){Gingell}, {Burgess}, \&
  {Matteini}}]{gingell:2015}
{Gingell}, P.~W., {Burgess}, D., \& {Matteini}, L. 2015, \apj, 802, 4

\bibitem[{{Gustafson} {et~al.}(2012){Gustafson}, {Ricci}, {Furno}, \&
  {Fasoli}}]{gustafson:2012}
{Gustafson}, K., {Ricci}, P., {Furno}, I., \& {Fasoli}, A. 2012, Physical
  Review Letters, 108, 035006

\bibitem[{{Hellinger} {et~al.}(2015){Hellinger}, {Matteini}, {Landi},
  {Verdini}, {Franci}, \& {Tr{\'a}vn{\'{\i}}{\v c}ek}}]{hellinger:2015}
{Hellinger}, P., {Matteini}, L., {Landi}, S., {et~al.} 2015, \apjl, 811, L32

\bibitem[{{Hietala} {et~al.}(2015){Hietala}, {Drake}, {Phan}, {Eastwood}, \&
  {McFadden}}]{hietala:2015}
{Hietala}, H., {Drake}, J.~F., {Phan}, T.~D., {Eastwood}, J.~P., \& {McFadden},
  J.~P. 2015, \grl, 42, 7239

\bibitem[{{Jokipii}(2013)}]{jokipii:2013}
{Jokipii}, J.~R. 2013, \ssr, 176, 115

\bibitem[{{K{\'o}ta} \& {Jokipii}(2008)}]{kota:2008}
{K{\'o}ta}, J., \& {Jokipii}, J.~R. 2008, AIP Conf. Series, 1039, 397

\bibitem[{{Krimigis} {et~al.}(2013){Krimigis}, {Decker}, {Roelof}, {Hill},
  {Armstrong}, {Gloeckler}, {Hamilton}, \& {Lanzerotti}}]{krimigis:2013}
{Krimigis}, S.~M., {Decker}, R.~B., {Roelof}, E.~C., {et~al.} 2013, Science,
  341, 144

\bibitem[{{Lazarian} \& {Opher}(2009)}]{lazarian:2009}
{Lazarian}, A., \& {Opher}, M. 2009, \apj, 703, 8

\bibitem[{{le Roux} {et~al.}(2015){le Roux}, {Zank}, {Webb}, \&
  {Khabarova}}]{leroux:2015}
{le Roux}, J.~A., {Zank}, G.~P., {Webb}, G.~M., \& {Khabarova}, O. 2015, \apj,
  801, 112

\bibitem[{{Matteini} {et~al.}(2013){Matteini}, {Landi}, {Velli}, \&
  {Matthaeus}}]{matteini:2013}
{Matteini}, L., {Landi}, S., {Velli}, M., \& {Matthaeus}, W.~H. 2013, \apj,
  763, 142

\bibitem[{{Matthews}(1994)}]{matthews:1994}
{Matthews}, A.~P. 1994, J. Comp. Phys., 112, 102

\bibitem[{{McComas} \& {Schwadron}(2006)}]{mccomas:2006}
{McComas}, D.~J., \& {Schwadron}, N.~A. 2006, \grl, 33, 4102

\bibitem[{{Metzler} \& {Klafter}(2000)}]{metzler:2000}
{Metzler}, R., \& {Klafter}, J. 2000, \physrep, 339, 1

\bibitem[{{Pommois} {et~al.}(2007){Pommois}, {Zimbardo}, \&
  {Veltri}}]{pommois:2007}
{Pommois}, P., {Zimbardo}, G., \& {Veltri}, P. 2007, Phys. Plasmas, 14, 012311

\bibitem[{{Richardson}(2011)}]{richardson:2011}
{Richardson}, J.~D. 2011, \apj, 740, 113

\bibitem[{{Schindler} {et~al.}(1988){Schindler}, {Hesse}, \&
  {Birn}}]{schindler:1988}
{Schindler}, K., {Hesse}, M., \& {Birn}, J. 1988, \jgr, 93, 5547

\bibitem[{{Schoeffler} {et~al.}(2011){Schoeffler}, {Drake}, \&
  {Swisdak}}]{schoeffler:2011}
{Schoeffler}, K.~M., {Drake}, J.~F., \& {Swisdak}, M. 2011, \apj, 743, 70

\bibitem[{{Senanayake} {et~al.}(2015){Senanayake}, {Florinski}, {Cummings}, \&
  {Stone}}]{senanayake:2015}
{Senanayake}, U.~K., {Florinski}, V., {Cummings}, A.~C., \& {Stone}, E.~C.
  2015, \apj, 804, 12

\bibitem[{{Stone} {et~al.}(2005){Stone}, {Cummings}, {McDonald}, {Heikkila},
  {Lal}, \& {Webber}}]{stone:2005}
{Stone}, E.~C., {Cummings}, A.~C., {McDonald}, F.~B., {et~al.} 2005, Science,
  309, 2017

\bibitem[{{Stone} {et~al.}(2008){Stone}, {Cummings}, {McDonald}, {Heikkila},
  {Lal}, \& {Webber}}]{stone:2008}
---. 2008, \nat, 454, 71

\bibitem[{{Swisdak} {et~al.}(2013){Swisdak}, {Drake}, \&
  {Opher}}]{swisdak:2013}
{Swisdak}, M., {Drake}, J.~F., \& {Opher}, M. 2013, \apjl, 774, L8

\bibitem[{{Vlahos} {et~al.}(2004){Vlahos}, {Isliker}, \&
  {Lepreti}}]{vlahos:2004}
{Vlahos}, L., {Isliker}, H., \& {Lepreti}, F. 2004, \apj, 608, 540

\bibitem[{{Wicks} {et~al.}(2010){Wicks}, {Horbury}, {Chen}, \&
  {Schekochihin}}]{wicks:2010}
{Wicks}, R.~T., {Horbury}, T.~S., {Chen}, C.~H.~K., \& {Schekochihin}, A.~A.
  2010, \mnras, 407, L31

\bibitem[{{Yeates} \& {Hornig}(2011)}]{yeates:2011}
{Yeates}, A.~R., \& {Hornig}, G. 2011, Phys. Plasmas, 18, 102118

\bibitem[{{Yoo} \& {Cho}(2014)}]{yoo:2014}
{Yoo}, H., \& {Cho}, J. 2014, \apj, 780, 99

\bibitem[{{Zank} {et~al.}(2014){Zank}, {le Roux}, {Webb}, {Dosch}, \&
  {Khabarova}}]{zank:2014}
{Zank}, G.~P., {le Roux}, J.~A., {Webb}, G.~M., {Dosch}, A., \& {Khabarova}, O.
  2014, \apj, 797, 28

\bibitem[{{Zank} {et~al.}(2015){Zank}, {Hunana}, {Mostafavi}, {Le Roux}, {Li},
  {Webb}, {Khabarova}, {Cummings}, {Stone}, \& {Decker}}]{zank:2015}
{Zank}, G.~P., {Hunana}, P., {Mostafavi}, P., {et~al.} 2015, \apj, 814, 137

\bibitem[{{Zimbardo} \& {Perri}(2013)}]{zimbardo:2013}
{Zimbardo}, G., \& {Perri}, S. 2013, \apj, 778, 35

\bibitem[{{Zimbardo} {et~al.}(2006){Zimbardo}, {Pommois}, \&
  {Veltri}}]{zimbardo:2006}
{Zimbardo}, G., {Pommois}, P., \& {Veltri}, P. 2006, \apjl, 639, L91

\end{thebibliography}

\clearpage

\begin{figure}
\begin{tabular}{cc}
(a) &\\[-1ex]
&\includegraphics[width=7cm]{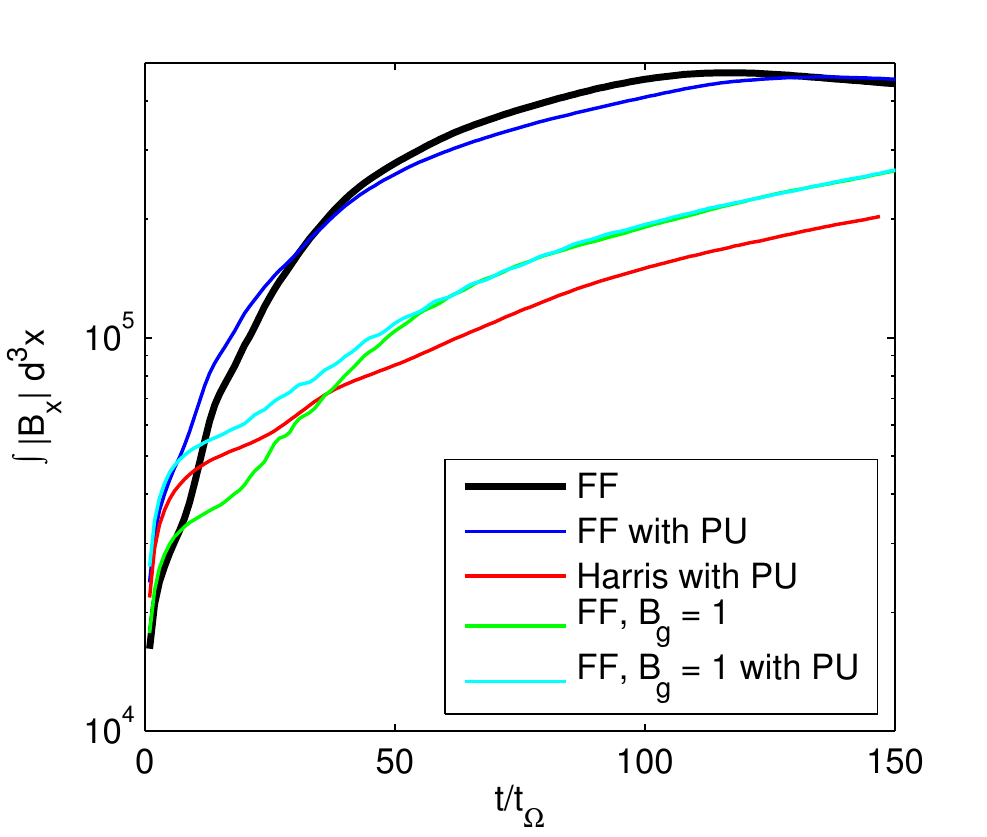}\\
(b) &\\[-1ex]
&\includegraphics[width=7cm]{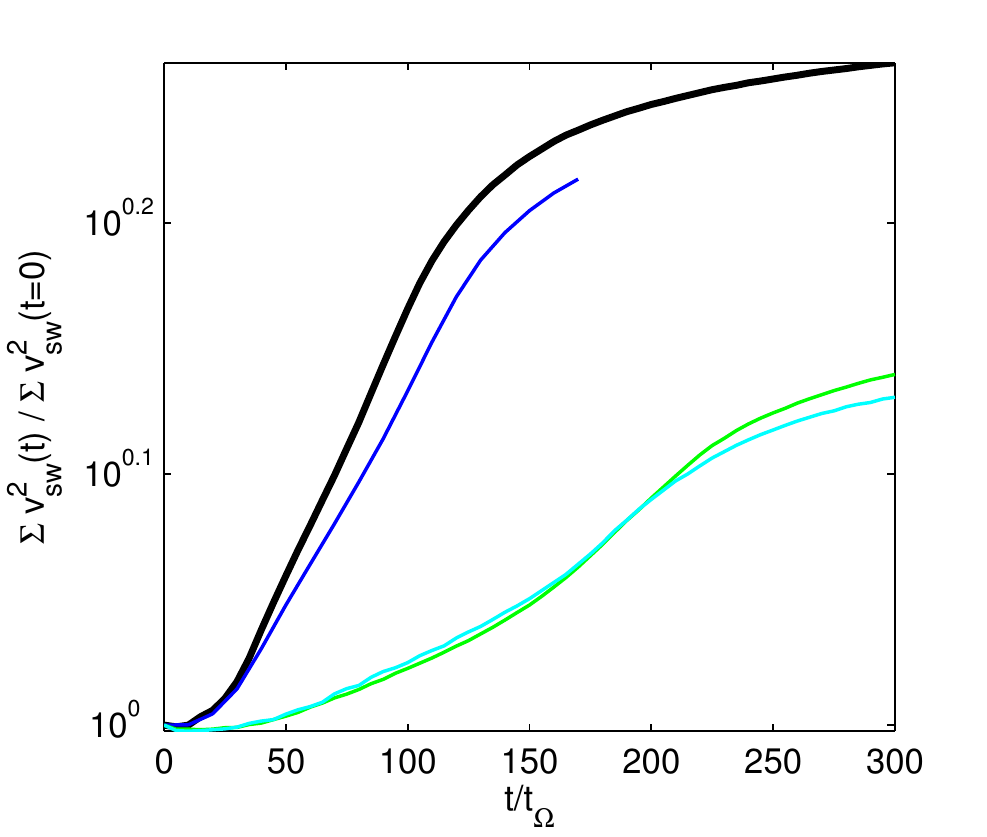}\\
\end{tabular}
\caption{Time evolutions of (a) $\int |B_x|d^3x$ (as a proxy of reconnected flux), and (b) total kinetic energy of SW population as a ratio to initial value. Results are shown for force-free Harris (FF) and Harris current sheets, with and without PUI, and with ($B_\mathrm{g}=1$) and without a guidefield.}
\label{fig:2}
\end{figure}

\begin{figure}
\includegraphics[width=16cm]{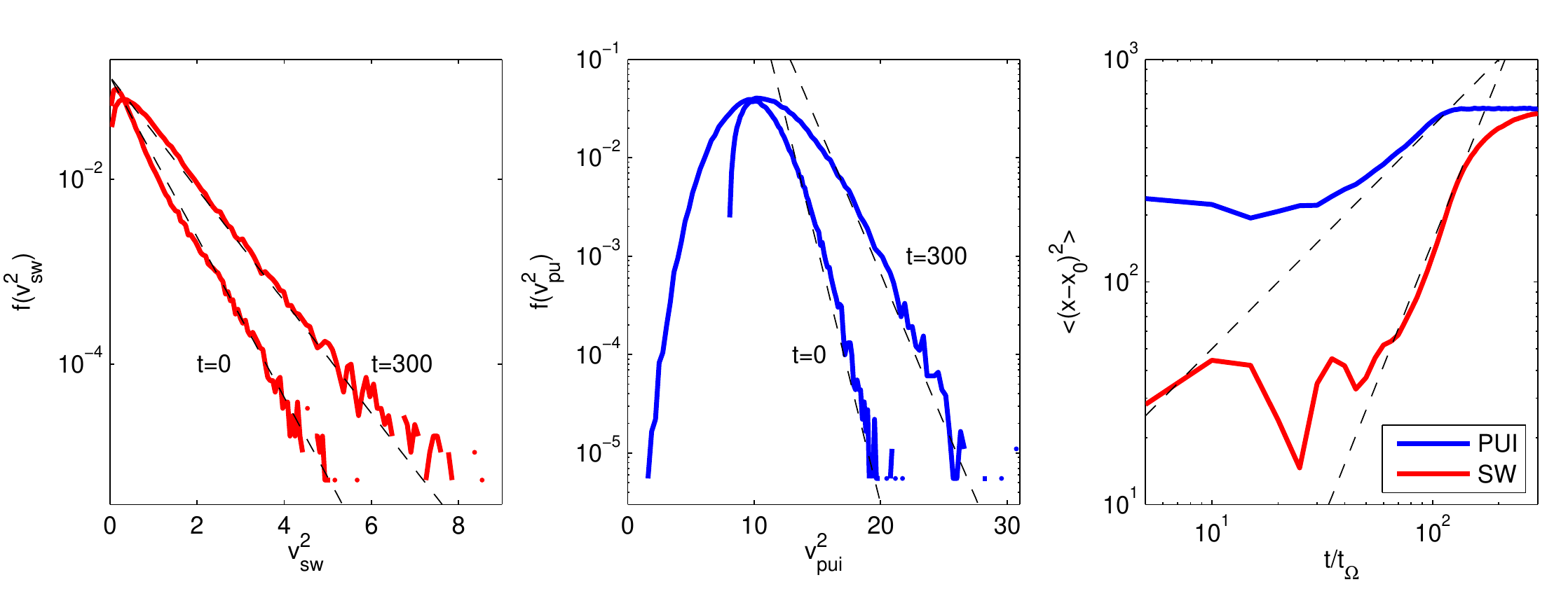}
\caption{Particle energy distributions for SW (left panel) and  PUI (middle panel) populations at $t=0$ and $t=300t_\Omega$. Time evolution of mean square displacement for solar wind (red) and PUI (blue) populations (right panel).}
\label{fig:3}
\end{figure}

\begin{figure}
\begin{tabular}{cc}
(a) &\\[-1ex]
& \includegraphics[width=7cm]{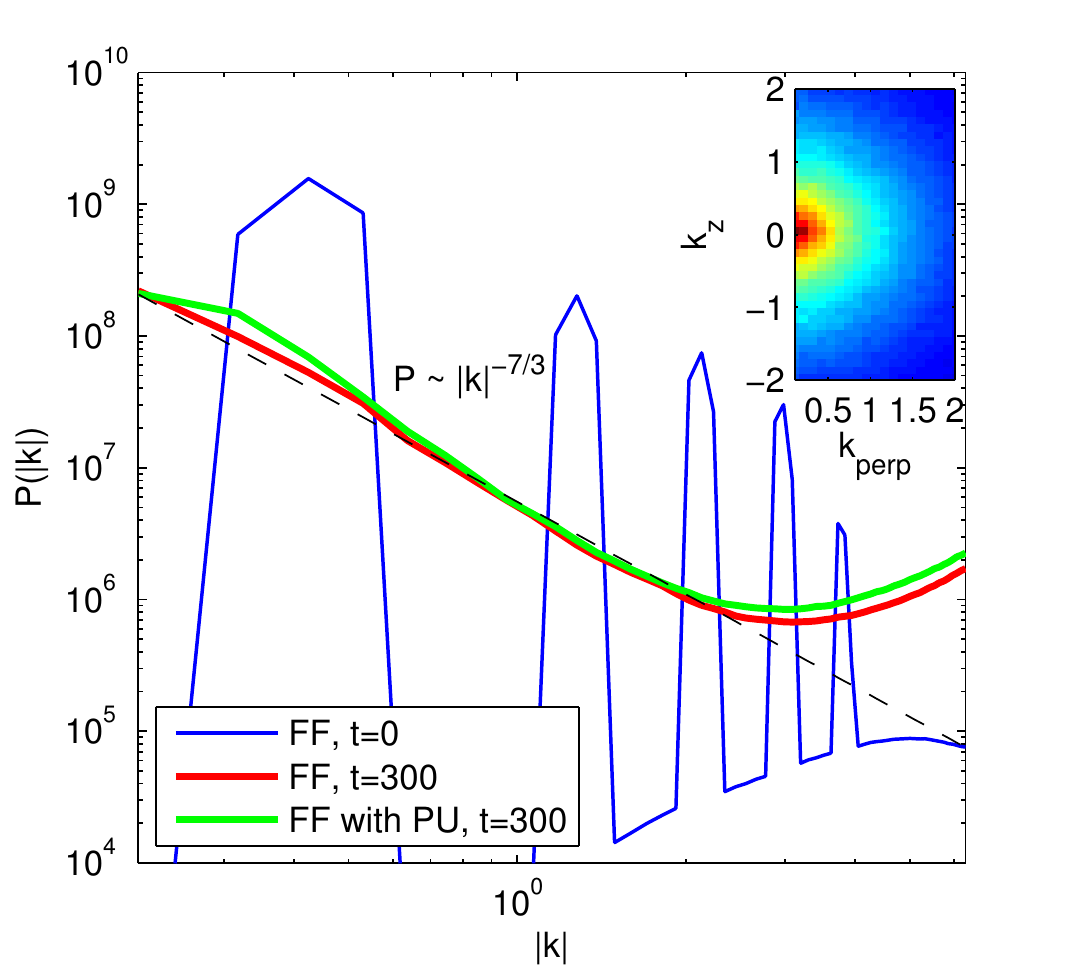}\\
(b) &\\[-1ex]
&\includegraphics[width=7cm]{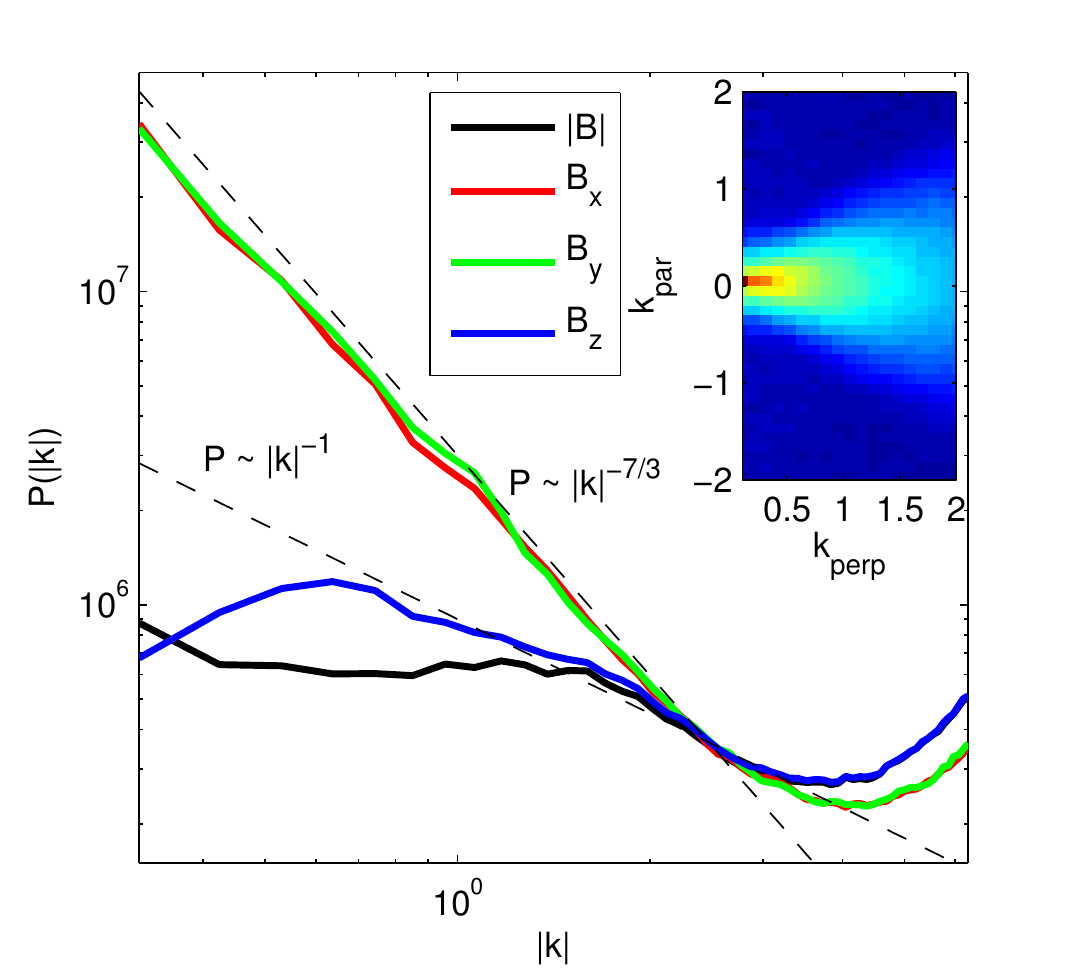}\\
\end{tabular}
\caption{Spectra of magnetic fluctuations at the end of the simulation for the force-free current sheet case, for (a) total power for zero guide field, and (b) total and component power for non-zero guide field. Insets for (a) and (b) show the power distribution as a function of $k_z - k_\perp$ and $k_\parallel - k_\perp$, respectively. Panel (a) also shows the spectrum at t=0.}
\label{fig:4}
\end{figure}

\begin{figure}
\includegraphics[width=16cm]{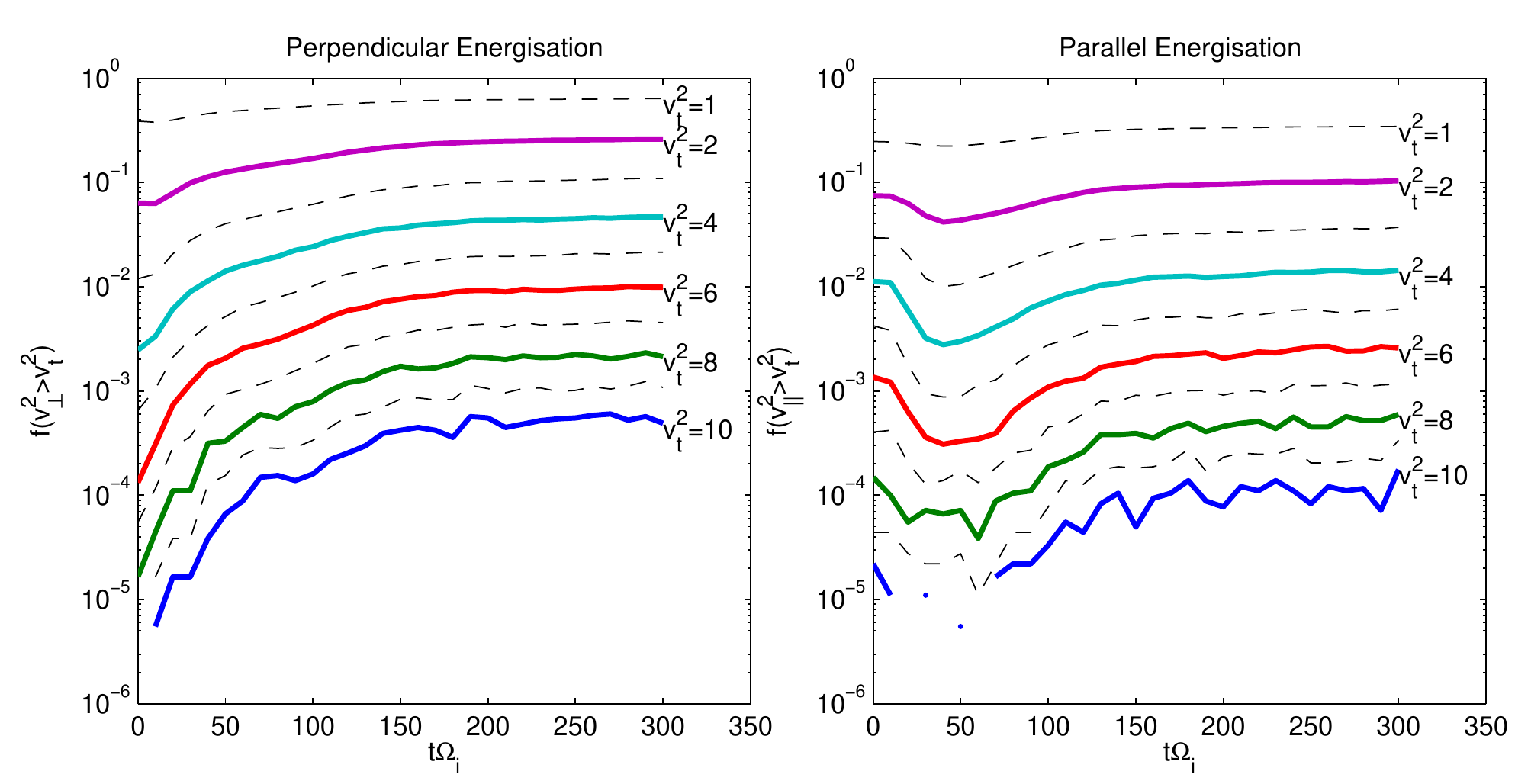}
\caption{Time evolution of the fraction of solar wind ions with perpendicular (left) and parallel (right) energy above a given threshold $v_t^2$ in units of $v_A^2$.}
\label{fig:5}
\end{figure}

\begin{figure}
\includegraphics[width=16cm]{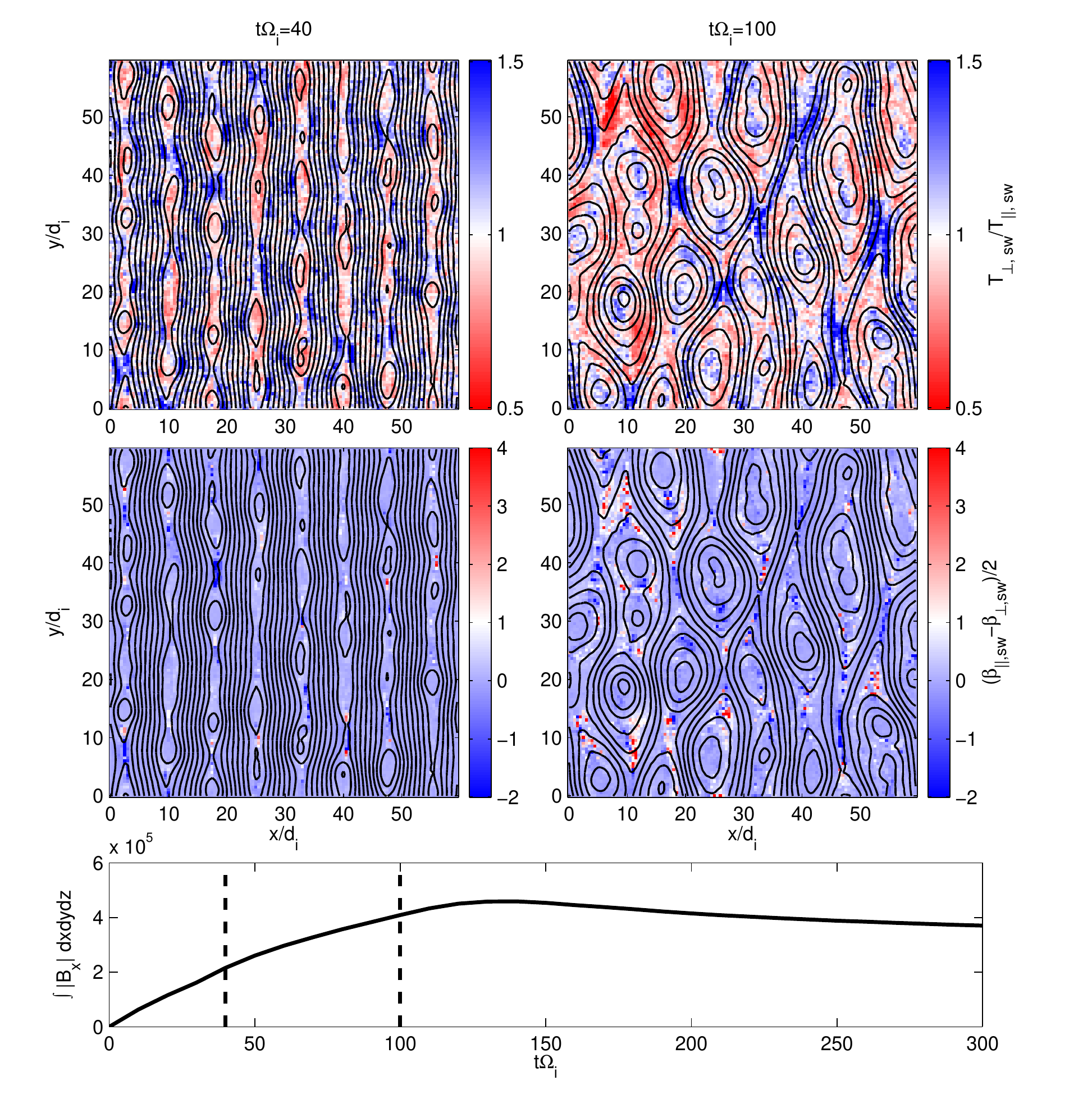}
\caption{Top row: temperature anisotropy $T_{\perp}/T_{\parallel}$ for the solar wind ion population during the linear (left) and non-linear (right) phases of the tearing instability. Middle row: Firehose instability threshold  $(\beta_{\parallel} -\beta_{\perp})/2$, showing stable (blue) and unstable (red) regions at $t\Omega_\mathrm{i} = 40,100$. Field lines are displayed in both rows in black. Bottom row: time evolution of the reconnected flux, as in Figure~2a, highlighting the times shown in  the left and right columns.}
\label{fig:6}
\end{figure}

\end{document}